\documentstyle[sprocl,epsfig]{article}

\newcommand \beq{\begin{eqnarray}}
\newcommand \eeq{\end{eqnarray}}
\def\del{\partial}                              

\def\frac#1#2{{#1 \over #2}}
\def\simge{\mathrel{%
   \rlap{\raise 0.511ex \hbox{$>$}}{\lower 0.511ex \hbox{$\sim$}}}}
\def\simle{\mathrel{
   \rlap{\raise 0.511ex \hbox{$<$}}{\lower 0.511ex \hbox{$\sim$}}}}

\def\journal#1#2#3#4{\ {#1}{\bf #2}, {#4} ({#3})}
\def\NPB{\journal{Nucl.\ Phys.\ {\bf B}}}

\def\PLB{\journal{Phys.\ Lett.\ {\bf B}}}

\def\PRD{\journal{Phys.\ Rev.\ {\bf D}}}
\def\PRL{\journal{Phys.\ Rev.\ Lett.}}

\def\be{\begin{equation}}
\def\ee{\end{equation}}
\def\bea{\begin{eqnarray}}
\def\eea{\end{eqnarray}}

\begin{document}

\begin{titlepage}
\vspace*{-1.5cm}
\begin{flushright} {Saclay-T98/073}
\end{flushright}
\vspace*{1.5cm}
\begin{center}
\baselineskip=13pt
{\large {\bf CLASSICAL EFFECTIVE THEORY FOR HOT QCD\\}}
\vskip0.5cm
Edmond Iancu\\
{\it Service de Physique Th\'eorique\footnote{Laboratoire de la Direction
des
Sciences de la Mati\`ere du Commissariat \`a l'Energie
Atomique}, CE-Saclay \\ 91191 Gif-sur-Yvette, France}\\

\vskip 2.cm
\end{center}

\begin{abstract}
In high temperature QCD, the perturbation theory is plagued with
infrared divergences which reflect long-range non-perturbative phenomena.
I argue that it is possible to study such phenomena within a {\it classical}
thermal field theory which can be put on a three-dimensional lattice. 
The classical theory is an effective theory for the soft, non-perturbative 
modes, as obtained after integrating out the hard modes in perturbation theory.
It is well suited for numerical studies of the non-perturbative real-time
dynamics, which cannot be studied within the standard, imaginary-time
formulations of lattice QCD.
\end{abstract}

\vspace*{5.cm}

\begin{flushleft}

Invited talk at the Third Workshop on ``Continuous Advances in QCD'',
University of Minnesota, Minneapolis, April 16--19, 1998.

\end{flushleft}

\end{titlepage}

\title{CLASSICAL EFFECTIVE THEORY FOR HOT QCD}

\author{E. IANCU}

\address{Service de Physique Th\'eorique, CEA Saclay \\
 91191 Gif-sur-Yvette, France\\E-mail: eiancu@cea.fr}

\maketitle\abstracts{
In high temperature QCD, the perturbation theory is plagued with
infrared divergences which reflect long-range non-perturbative phenomena.
I argue that it is possible to study such phenomena within a {\it classical}
thermal field theory which can be put on a three-dimensional lattice. 
The classical theory is an effective theory for the soft, non-perturbative 
modes, as obtained after integrating out the hard modes in perturbation theory.
It is well suited for numerical studies of the non-perturbative real-time
dynamics, which cannot be studied within the standard, imaginary-time
formulations of lattice QCD.}

\section{Introduction}

There is currently a considerable interest in the
physics of {\it ultrarelativistic plasmas}, by which we mean plasmas in,
or near, thermal equilibrium at very high temperature, much larger
than any mass scale: $T \gg m$. This interest, which has 
triggered many theoretical advances, is mainly motivated by 
two important applications: The first is to 
the deconfined phase of QCD, the celebrated quark-gluon plasma (QGP), which 
is expected to be found in the heavy ion collisions at RHIC and LHC.
The second is to the high-temperature, symmetric, phase of the electroweak
theory, which is relevant for baryogenesis in the early Universe. 

If the temperature is high enough, the physics of non-Abelian gauge
theories is expected to be simple. This common wisdom, which we shall see 
below to be a little too optimistic, is motivated by asymptotic freedom:
as the temperature increases,
the coupling ``constant'' becomes small, $g(T) \ll 1$,
and perturbation theory becomes applicable. 
That is, the high-$T$ plasma can be treated as
a gas of weakly interacting quarks and gluons \footnote{For definitness, 
I shall mostly use a QCD-inspired terminology. Note, however, that most
of the present considerations do also apply to the electroweak plasma,
provided the temperature is high enough ($T\gg T_{c}$).}.

By using perturbation theory, significant progress 
has been indeed achieved in understanding the long-wavelength
excitations of the plasma and the related screening phenomena 
(see Refs. \cite{BIO96,MLB} for a summary and more references,
and Sec. 2 below for a brief account). Furthermore, the free energy
in hot QCD has been computed \cite{AZ} up to the order $g^5$,
which is the highest accuracy permitted by perturbation theory
(the correction $O(g^6)$ turns out to be non-perturbative; see Sec. 
3 below).

At the same time, lattice simulations have been used
to compute thermodynamic quantities \cite{DeTar}
 (like the free energy and the entropy)
and to study static aspects of the deconfining \cite{DeTar}
and the electroweak \cite{Kajantie} phase transitions.
Besides providing ``exact'' numerical results, the lattice calculations 
allow us to explore non-perturbative physics like the strong coupling
regime in QCD at intermediate temperatures
($T \simge T_{c} \sim 200$ MeV). As the temperature increases,
the comparaison with perturbation theory becomes possible:
it has been found, for instance, that
the lattice estimates for the energy density
approach rather quickly the Stefan-Boltzmann limit above $T_{c}$,
thus comforting the picture of hot QCD as a weakly interacting gas
\footnote{Rather surprisingly, it turns out that the perturbative expansion
for the free energy \cite{AZ} is poorly convergent up to
temperatures as high as 3 GeV. Still, as shown in Refs. \cite{Hatsuda},
the convergence can be greatly improved by using Pad\'e approximants.}.

It thus may come as a surprise that, in some cases, lattice simulations
cannot be avoided not even in the study of the high temperature limit,
where the coupling constant is arbitrarily small. This is so since
perturbation theory is plagued with infrared
divergences which reflect large collective effects (cf. Sec. 3).
Moreover, such divergences affect not only static characteristics,
like the free energy, but also dynamical quantities, that is,
{\it real-time} correlation functions, for which the standard lattice
simulations --- as formulated in imaginary time --- are not applicable.
An important example, to which I shall return later,
is the rate of baryon number violation at high temperature \cite{Shapo}. To
compute such quantities, we need new, non-perturbative, methods which
allow for numerical studies of the real-time dynamics.
It is my purpose in this talk to present such a method which
is based upon a semiclassical approximation (cf. Sec. 4).

\section{Collective behaviour and screening}

I consider a purely Yang-Mills plasma \footnote{At high-$T$, quarks
are not important for the infrared physics to be discussed here.}
in thermal equilibrium at a temperature $T$ which is high enough 
for the coupling constant to be small: $g\ll 1$.
In the absence of interactions, this would be simply the
``black body radiation'', that is, a collection of free,
massless gluons with typical energies of the order $T$
and Bose-Einstein occupation numbers:
$N_0(E)=1/({\rm e}^{\beta E}\,-\,1)$, with $\beta\equiv 1/T$. 
However, the gauge interactions --- although weak --- do significantly
change this picture. They give rise to an hierarchy of scales:
\beq
T\,\gg\,gT\,\gg\,g^2 T\,\gg\, g^4 T\,\cdots\, ,
\eeq
with the various scales corresponding to different physical phenomena.

Thus, the typical excitations of the plasma are ``hard'' gluons,
with momenta $k\sim T$. Such gluons can develop a collective
behaviour over a typical space-time scale $\sim 1/gT$, which is large
as compared to the mean interparticle distance $\sim 1/T$.
This results in long-wavelength ($\lambda \sim 1/gT \gg 1/T$) 
oscillations of the averge colour density which are most 
economically described in terms of {kinetic equations} \cite{BIO96}.
In these equations, the {\it hard} ($k \sim T$) gluons are represented by
their average colour density $\delta N_a({{\bf k}},x)$
to which couple the {\it soft} (i.e., long-wavelength)
colour fields $A_a^\mu(x)$. The relevant equations read:
\beq\label{N}
(D_\nu F^{\nu\mu})_a(x)&=&
2gC_A\int\frac{{\rm d}^3k}{(2\pi)^3}\,\,v^\mu
\,\delta N_a({\bf k},x),\nonumber\\
(v\cdot D_x)_{ab}\delta N^b({{\bf k}},x)&=&-\, g\,
{\bf v}\cdot{\bf E}_a(x)\,\frac{{\rm d}N_0}{{\rm d}k}\,,\eeq
where $D^\mu=\del^\mu+igA^\mu_a T_a$ is the covariant derivative,
$E_a^i\equiv F_a^{i0}$ is the chromoelectric field,
$C_A=N$ for SU($N$), and $v^\mu\equiv (1,\,{\bf v})$ with ${\bf v}=
{\bf k}/k$ denoting the velocity of the hard particles ($k= |{\bf k}|$,
so that $|{\bf v}| =1$). 

The first equation above is the Yang-Mills equation for the soft fields
$A_a^\mu$. It involves, in its r.h.s., the  colour current 
induced by the collective motion of the hard particles:
\beq\label{j}
j^{\mu}_{a}(x)
\equiv 2gC_A\int\frac{{\rm d}^3k}{(2\pi)^3}\,v^\mu
\,\delta N_a({\bf k},x).\eeq
In turn, the soft colour wave $A_a^\mu(x)$ acts as a driving force for
the collective behaviour, and this is described by the second Eq.~(\ref{N})
(which may be seen as a non-Abelian generalization of the familiar
Vlasov equation \cite{BIO96,Kelly}).  By solving this equation,
we can express the current $j_a^\mu$ in terms of the gauge fields $A_a^\mu$
and thus obtain an {\it effective} Yang-Mills equation which involves 
the soft fields alone:
\beq\label{ava}
D_\nu F^{\nu\mu}\,=\,m_D^2\int\frac{{\rm d}\Omega}{4\pi}\,
\frac{v^\mu v^i}{v\cdot D}\,E^i.\eeq
Here, the angular integral $\int {\rm d}\Omega$ runs over the 
orientations of the unit vector ${\bf v}$, and $m_D$ is the {\it Debye mass\,}:
\beq\label{MD}
m^2_D\equiv -\frac{g^2 C_A}{\pi^2}\int_0^\infty 
{\rm d}k \,k^2\,\frac{{\rm d}N_0}{{\rm d}k}\,=\,\frac{g^2 C_A T^2}{3}\,.
\eeq
Eq.~(\ref{ava}) describes the propagation of long-wavelength colour waves
in the high-$T$ plasma. The induced current in its r.h.s. is the result
of the wave scattering off the hard thermal particles.
In general, this current is non-local, and also non-linear in the
gauge fields (note the covariant derivative in the denominator).
Still, for {\it time-independent} fields, it reduces to a very simple
expression: $j^\mu_a({\bf x})=\delta^{\mu 0}m_D^2 A^0_a ({\bf x})$,
which defines a screening ``mass'' for the Coulomb field $A^0_a$.
Indeed, for such static fields, the $\mu = 0$ component of  Eq.~(\ref{ava})
simplifies to:
\beq\label{PD}
{\bf D\cdot E} + m_D^2 A_0({\bf x}) = 0,\eeq
which is the non-Abelian generalization of the Poisson-Debye equation:
\beq (-\Delta + m_D^2)  A_0({\bf x}) = 0,\eeq
and implies the screening of any electrostatic fluctuation
$A_0({\bf x})$ over distances $r \sim 1/m_D$.
In other terms, the  static ($k_0\to 0$) limit of the
Coulomb propagator --- as following from Eq.~(\ref{PD}) --- reads:
\beq\label{C}
D_{00}(k_0=0,{\bf k})\,=\,\frac{1}{{\bf k}^2 + m_D^2}\,,\eeq
so that the Debye mass acts as an infrared cutoff $\sim gT$
in the electric sector.

The situation in the magnetic sector is more complex: in the static
limit the vector current ${\bf j}_a$ vanishes, as alluded to before,
so that the time-independent magnetic fields are not screened. 
The analogue of Eq.~(\ref{C}) reads then:
\beq\label{M}
D_{ij}(k_0=0,{\bf k})\,=\,\frac{\delta_{ij}-\hat k_i \hat k_j}
{{\bf k}^2}\,,\eeq
which is the same as at tree level.
For {\it time-dependent} magnetic fields, however,
screening does occur, through the mechanism of {\it Landau damping}.
This too can be studied \cite{BIO96,MLB} on Eq.~(\ref{ava}),
with the result that, for small but non-vanishing frequency $k_0$,
the magnetic piece of the gluon propagator reads:
\beq\label{MT}
D_{ij}(k_0, {\bf k})\,\simeq\,\frac{\delta_{ij}-\hat k_i \hat k_j}
{ k^2\,-\,i(\pi k_0/4k)m_D^2}\,.\eeq
Note the purely imaginary character of the self-energy in the
denominator: this is a dissipative effect, describing the absorbtion
of the magnetic field by the plasma constituents. Eq.~(\ref{MT}) also
shows that, for large enough frequencies $k_0\sim k$,
the magnetic fields are screened as efficiently as the electric ones.

The above kinetic-theory picture of the screening (which is 
actually equivalent to the one-loop picture in the
``hard thermal loop'' approximation \cite{BIO96,MLB}) turns out
to be further complicated by non-perturbative phenomena.
There are indeed theoretical arguments \cite{MLB}, which are also
supported by lattice simulations \cite{DeTar}, and which suggest that
screening should occur also for the {\rm static} magnetic fields,
but only at the softer scale $g^2 T$ (see Sec. 4 below).
In practical calculations,
this is often parametrized by introducing an infrared cutoff
$\mu \sim g^2 T$ in the magnetic sector (``magnetic mass'').
But it is fair to say that the corresponding physical
mechanism is not yet fully understood. This is so since,
as we shall see shortly, $g^2 T$ is
precisely the scale where perturbation theory breaks down.

\section{The breakdown of the perturbation theory}

The screening phenomena greatly improve
the infrared behaviour of the perturbation theory (PT), thus allowing
for many interesting calculations \cite{MLB}. Still, when
going to higher orders in PT, one is often confronted with severe
infrared (IR) divergences, which signal a breakdown of
the perturbation theory at the scale $g^2 T$. There are 
essentially two reasons for that: 

The first is the singularity in the
magnetostatic propagator as $k \equiv |{\bf k}| \to 0$ (cf. Eq.~(\ref{M})).
Although this might be screened, as alluded to before, at the scale $g^2 T$, 
nevertheless such a screening will not be enough to restore 
perturbation theory (see below). 

The second reason is the Bose-Einstein
amplification of the soft modes with momenta $k\ll T$:
such modes have large thermal occupation numbers,
\beq\label{BE}
N_0(k)\,\equiv\,\frac{1}{{\rm e}^{\beta k}-1}\,\simeq\,\frac{T}{k}
\,\,\gg \,1,\eeq
and therefore give large ``radiative'' corrections in higher orders.
Specifically, by adding a new loop to a preexistent Feynman graph, 
we generate a correction of relative order $g^2 N_0(k)$, where $k$ 
is the momentum carried by the added gluon propagator. 
If $k\sim g^2 T$, then $N_0(k)\sim T/k \sim 1/g^2$ and $g^2 N_0(k) \sim 1\,$: 
that is, by adding more soft ($k\sim g^2T$) loops, we remain at the same
order in $g$, and the loop expansion breaks down.

To see a specific example of this difficulty, let's follow Linde
\cite{MLB} and consider the higher order corrections to the free energy:
\begin{figure}
\protect \epsfysize=2.5cm{\centerline{\epsfbox{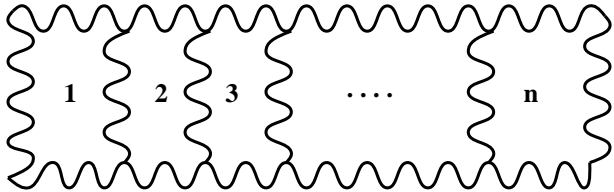}}}
	 \caption{$n$-loop ladder diagram contributing to the free energy.}
\label{ladder}
\end{figure}
a typical $n$-loop (with $n\ge 4$) diagram has
the ``ladder'' topology in Fig.~\ref{ladder} and gives a contribution
$F^{(n)}\sim g^6T^4\left(g^2T/\mu\right)^{n-4}$
(this is a simple power counting estimate).
In this equation, $\mu$ is an infrared cutoff which has been
introduced by hand to give a meaning to an otherwise IR-divergent
loop integral. As we know by now, such a cutoff is indeed generated
via the screening effects. In the {\it electric} sector, we have Debye
screening and therefore $\mu\sim m_D \sim gT$ (cf. Eq.~(\ref{C}));
with $\mu \sim gT$, $F^{(n)}\,\sim\,g^{n+2}T^4$, and higher loops contribute 
to higher orders in $g$, as it should for PT to make sense.
In the {\it magnetic} sector, on the other hand, we have at most $\mu \sim g^2T$, 
in which case all the diagrams with four or more loops
contribute to the {\it same} order in $g$ (namely, to the order $g^6$).
We thus face a breakdown of PT in the {magnetic} sector
which, unlike the {electric} sector, is not protected by Debye screening.

For {\it static} quantities like the free energy, the non-perturbative
corrections can be estimated, at least in principle, via lattice QCD.
But there are also {\it time-dependent} correlation functions which appear to be
non-perturbative and for which the standard lattice calculations (as formulated
in imaginary time) are not applicable.
Let me give you some examples in this sense:
 
In relation to baryogenesis, one is interested
in the rate for anomalous baryon number violation in the
high-$T$, symmetric, phase of the electroweak theory \cite{Shapo}.
This is a genuinely non-perturbative phenomenon where the variation
$\Delta B$ of the baryon number is tied up --- via the chiral anomaly --- 
to the transitions between topologically inequivalent vacua:
\beq\label{B}
\Delta B (t)\propto \int_0^t {\rm d}x_0\int {\rm d}
^3x \,F^a_{\mu\nu} {\tilde F}_a^{\mu\nu},\eeq
with ${\tilde F}_a^{\mu\nu} \equiv (1/2)\varepsilon^{\mu\nu\rho\lambda} 
F^a_{\rho\lambda}$. At high temperature, such transitions
will necessarily involve very soft ($k \sim g^2 T$) magnetic fields, the only
ones to be responsible for non-perturbative phenomena.
Still, it has been recently recognized \cite{Yaffe} that the
topological transitions are also sensitive to the {\it hard} ($k\sim T$) 
plasma modes, via the Landau damping alluded to before (cf. Eq.~(\ref{MT})).

\begin{figure}
\protect \epsfysize=3.cm{\centerline{\epsfbox{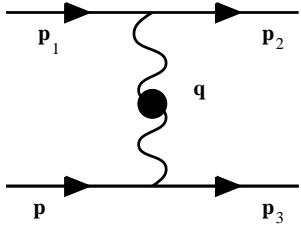}}}
	 \caption{Fermion-fermion elastic scattering in the Born approximation.
The bubble on the gluon line indicates the resummation of the screening
effects.}
\label{Born}
\end{figure}
But non-perturbative effects are also met in the study of simpler correlation
functions, like the gluon, or quark, propagator: e.g., when studying
the spectrum of the elementary excitations (the plasma ``quasiparticles''),
we consider the large time behaviour of the corresponding 2-point functions.
On general grounds, one expects an {\it exponential} decay of the quasiparticles,
as coming from their scattering off the plasma constituents. For instance, if
$S(t,{\bf p})=\,<\bar\psi(t,{\bf p})\psi(t,-{\bf p})>$ denotes the 
quark propagataor, then one expects $S(t\to \infty,{\bf p})
\propto {\rm e}^{iE(p) t}\,{\rm e}^{-\gamma(p)\,t}$,
where $E(p)$ is the mass-shell energy and $\gamma(p)$ is the {damping rate},
i.e., the total interaction rate of the quark in the plasma.
Still, explicit calculations \cite{Pisarski93}
of $\gamma(p)$ show a logarithmic infrared 
divergence already in leading order, as coming from collisions with
the exchange of soft magnetic gluons (cf.  Fig. \ref{Born}).
Similar divergences occur for gluons, and also for electrons
in a hot QED plasma. Thus, the lifetime of the quasiparticles
turns out not to be computable in perturbation theory.


\section{Classical effective theory for real-time processes}

I now present a method \cite{Iancu} which allows us, at least in principle,
to compute the non-perturbative, real-time, correlation functions
alluded to before.

The basic idea is not new  \cite{Shapo}:  since the
non-perturbative phenomena are associated with soft
($k \sim g^2T$) magnetic fields which have large thermal occupation numbers
(cf. Eq.~(\ref{BE})), some semi-classical approximation should be applicable.
To see this in a simple way, let's put back Planck's constant $\hbar$
in Eq.~(\ref{BE}) and compute the average energy per mode in thermal equilibrium:
\beq\label{EQP}
\varepsilon(k)\,=\,\frac{\hbar k}{{\rm e}^{\beta \hbar k}-1}\,\simeq\,\,{T}
\qquad{\rm as}\,\,\,\, \,\,\hbar k \ll T.
\eeq
As $\hbar \to 0$, we recover the classical equipartition theorem,
as expected. But the above example shows that the relevant inequality
is $\hbar k \ll T$, so that the classical limit ($\hbar \to 0$ at fixed
$k$ and $T$) is actually equivalent to the soft momentum limit
($k \to 0$ at fixed $\hbar$ and $T$).
This observation is useful since we know how to perform real-time lattice 
simulations for a {\it classical} thermal field theory: 
All we have to do is to solve the classical equations of motion for given 
initial conditions, and then average over the classical phase
space with the Boltzmann weight exp$(-\beta H)$. Since the initial
conditions (say $\phi({\bf x})$ and $\dot\phi({\bf x})$ for a scalar theory)
depend only on the spatial coordinate ${\bf x}$, the phase space integration
is actually a {\it three-dimensional} functional integral, which can be
implemented on a lattice in the standard way (and actually, with even less
numerical effort than in four dimensions !). The only question is,
what is the correct classical theory ?

It has been originally assumed \cite{Shapo} that, in order
to compute the hot baryon number violation 
(cf. Eq.~(\ref{B})), it should be enough to consider
the classical Yang-Mills theory at finite temperature.
This hypothesis, which led to the lattice calculations in Ref. \cite{AmbK},
is highly non-trivial: it assumes that the topological
transitions are totally insensitive to the hard ($k\sim T$) plasma
modes, which are not properly described by the classical theory
(since the approximation in Eq.~(\ref{EQP}) fails at momenta
$k\sim T$). And indeed, the classical theory is well-known
to run into ultraviolet (UV) problems, like the famous
``ultraviolet catastrophe'' of Rayleigh and Jeans:
The classical estimate for the energy density of the black body radiation,
namely ($\Lambda$ is an ad-hoc UV cutoff):
\beq E_{cl}/V\,=\,
\int\frac{{\rm d}^3k}{(2\pi)^3}\,\varepsilon_{cl}(k)\,=\,T
\int\frac{{\rm d}^3k}{(2\pi)^3}\,\propto\,T \Lambda^3,\eeq
is obviously wrong (since UV divergent), in contrast to the quantum result:
\beq E/V\,=\,
\int\frac{{\rm d}^3k}{(2\pi)^3}\,
\frac{k}{{\rm e}^{\beta k}-1}\,\propto\,{T^4},\eeq
which is finite since the large momenta $k\gg T$ are exponentially 
suppressed by the Bose-Einstein distribution function.

Now, as already argued in Sec. 3, the topological transitions in hot QCD
are driven by soft ($k\sim g^2 T$) field configurations,
and this is the main justification for using the classical Yang-Mills theory
in Refs. \cite{Shapo,AmbK}. Still, it has been also argued in Sec. 2 that
the dynamics of the soft modes is strongly modified by the hard particles, 
which generate screening. Thus, even though a soft process, the baryon
number violation might still be sensitive to the hard particles,
via the screening effects. And actually we have both theoretical
\cite{Yaffe} and numerical \cite{Hu} evidence that this is indeed the case.

Thus, in order to properly compute the
non-perturbative correlation functions of interest, we need to correct
the classical Yang-Mills theory by including the screening effects.
That is, the relevant classical theory should be an {\it effective} theory 
which applies only to the soft modes, but where the hard modes have been
integrated over to generate screening.
From Sec. 2, we have a candidate for such a theory: Eq.~(\ref{ava})
includes indeed the screening effects, via the colour current in its r.h.s.
Still, there are a few ``technical'' complications associated with
this equation: First, Eq.~(\ref{ava}) is non-local (and also dissipative:
recall the imaginary part in the denominator of Eq.~(\ref{MT})),
so it is not a priori clear how to construct the thermal phase-space
and the Hamiltonian. Second, in order to avoid overcounting,
we need a precise, and gauge-invariant, separation between hard and soft 
degrees of freedom. 

The discussion in Sec. 2 suggests a solution \cite{McLerran}
to the first problem above: rather than working with the non-local equation
\ref{ava}, we can conveniently replace it with the coupled system of {\it local}
equations in Eq.~(\ref{N}). There is a price to be payed for that:
in addition to the gauge fields $A^\mu_a(x)$, the local description
in Eq.~(\ref{N}) also involves the average colour density $\delta N_a({{\bf k}},x)$,
which can be seen as an ``auxiliary field''.
Still, when working with a local theory, we are in a better position
to look for a Hamiltonian formulation, as I discuss now.

The first step is to recognize, on the second Eq.~(\ref{N}),
that the ${\bf v}$ and $k$-dependence 
can be factorized in $\delta N^a({{\bf k}},x)$ by writing:
\beq\label{dn}
\delta N^a({\bf k}, x)\equiv - gW^a(x,{\bf v})\,({\rm d}N_0/{\rm d}k).\eeq
The new functions $W^a(x,{\bf v})$ satisfy the equation:
\beq\label{W}
(v\cdot D_x)_{ab}W^b(x,{\bf v})\,=\,{\bf v}\cdot{\bf E}_a(x),\eeq
which is independent of $k$ since the hard particles move at the speed
of light: $|{\bf v}| =1$. 
By using Eq.~(\ref{dn}), the induced current can be written as:
\beq\label{j1}
j^\mu_a(x)&=&m_D^2\int\frac{{\rm d}\Omega}{4\pi}
\,v^\mu\,W_a(x,{\bf v}),\eeq
where the radial integration (i.e., the integration over 
$k\equiv |{\bf k}|$) has been explicitly worked out, and
the Debye mass $m_D$ is defined in Eq.~(\ref{MD}).

The Hamiltonian formulation of the effective theory \cite{Nair,Iancu} involves
the auxiliary fields $W_a(x,{\bf v})$ together with the soft gauge fields
$A^\mu_a(x)$. In the temporal gauge $A^a_0=0$,
the independent degrees of freedom are 
$E^a_i$, $A^a_i$ and  $W^a$, and the corresponding equations of motion read
\footnote{Note that Eqs.~(\ref{CAN}) are not in canonical form: this is already
obvious from the fact that we have an odd number of equations.}:
\beq\label{CAN}
E^a_i&=&-\del_0 A^a_i,\nonumber\\
-\del_0 E^a_i +\epsilon_{ijk}(D_j B_k)^a &=&
m_D^2\int\frac{{\rm d}\Omega}{4\pi}\,v_i\,W^a(x,{\bf v}),\nonumber\\
\left(\del_0 + {\bf v\cdot D}\right)^{ab} W_b&=&{\bf v \cdot E}^a,\eeq
together with Gauss' law which in this gauge must be imposed as a constraint:
\beq\label{GAUSS}
G^a({\bf x})\equiv
({\bf D\cdot E})^a\,+\,m_D^2\int\frac{{\rm d}\Omega}{4\pi}\,W^a(x,{\bf v})
\,=\,0.
\eeq
Eqs.~(\ref{CAN}) are conservative; the corresponding, conserved
energy functional (which also acts as a Hamiltonian \cite{Iancu}) has the
following simple form \cite{BIO96}:
\beq\label{H}
H\,=\,\frac{1}{2}\int {\rm d}^3 x\left\{{\bf E}_a\cdot{\bf E}_a\,+\,
{\bf B}_a\cdot{\bf B}_a\,+\,m_D^2
\int\frac{{\rm d}\Omega}{4\pi}\,W_a(x, {\bf v})\,W_a(x, {\bf v})\right\},\eeq
which is manifestly gauge invariant.

We are now in position
to write down the classical partition function and compute
(generally time-dependent) thermal expectation values.
As discussed at the end of Sec. 3, we are interested in
correlation functions of the magnetic fields $A^i_a$. These can be
obtained from the following generating functional:
\beq\label{Z}
Z_{cl}[J^a_i]\,=\,
\int {\cal D}{\cal E}^a_i\,{\cal D}{\cal A}^a_i\,{\cal D}{\cal W}^a\,
\delta({\cal G}^a)\,
\exp\left\{-\beta {\cal H}\,+\,\int{\rm d}^4x \,J^a_i(x) A^a_i(x)\right\},\eeq
where $A^i_a(x)$ is the solution to Eqs.~(\ref{CAN})
with the initial conditions $\{{\cal E}^a_i,{\cal A}^a_i,{\cal W}^a\}$
(that is, $E^a_i(t_0,{\bf x})={\cal E}^a_i({\bf x})$, etc., with arbitrary $t_0$),
and  ${\cal G}^a$ and ${\cal H}$
are expressed in terms of the initial fields
(cf. Eqs.~(\ref{GAUSS}) and (\ref{H})).

It can be verified \cite{Iancu} that the phase-space measure 
${\cal D}{\cal E}^a_i{\cal D}{\cal A}^a_i{\cal D}{\cal W}^a$ in Eq.~(\ref{Z})
is invariant under the time evolution described by eqs.~(\ref{CAN}),
so that $Z_{cl}[J]$ is independent of the (arbitrary)
initial time $t_0$, as it should. (This point is not trivial because of the
non-canonical structure of the equations of motion.)

There is another essential --- but technically quite involved
 --- point that I am currently
glossing over: this is the intermediate cutoff separating hard from soft degrees
of freedom, and which should appear as an ultraviolet cutoff in Eq.~(\ref{Z})
(without such a cutoff, the effective theory would develop linear UV divergences
to one loop order).
As discussed in Ref. \cite{Iancu}, this cutoff can be indeed introduced
in such a way to make the effective theory UV finite, but cancel 
--- in the calculation of physical quantities --- against appropriate
``counterterms'' in the Hamiltonian. Moreover, the cutoff procedure proposed
in  Ref. \cite{Iancu} can be also implemented on a lattice; this is
important since it allows one to take the continuum limit in the lattice
calculations.

For illustration, let me finally consider two simple,
yet non-trivial, applications of the effective theory.
The first is the $J^a_i=0$ limit of  Eq.~(\ref{Z}) which yields, after some
simple algebra,
\beq\label{ZRED}
Z_{cl}=\int {\cal D}{\cal A}^a_0\,{\cal D}{\cal A}^a_i\,
\exp\left\{-\frac{\beta }{2}\int{\rm d}^3x \,\Bigl(
{\cal B}^a_i{\cal B}^a_i + ({\cal D}_i {\cal A}_0)^2
+ m_D^2{\cal A}_0^a{\cal A}_0^a
 \Bigr)\right\},\eeq
where the ${\cal A}_0^a$ components of the gauge fields have been reintroduced
as Lagrange multipliers to enforce Gauss' law.
This is ``almost'' the thermal partition function for classical Yang-Mills
theory: the only new feature is the screening mass
for the electric fields, which is the only
trace of the screening effects in the static limit (cf. Eq.~(\ref{PD})).

The expression in Eq.~(\ref{ZRED}) also coincides with the first order
result of the ``dimensional reduction'' \cite{Kajantie,Braaten}, a method
which consists in integrating out the non-static Matsubara modes \cite{MLB}
to obtain an effective 3-dimensional theory for the static one. 
By putting this theory on a lattice, one has been able to perform accurate
studies of the electroweak phase transition \cite{Kajantie} and also
to compute the non-perturbative contributions to the free-energy \cite{Karsch}
(cf. Sec. 3) and to the non-Abelian Debye mass \cite{Kajantie2}.

Note also that the magnetic sector of Eq.~(\ref{ZRED}) (that is,
the sector involving only the vector fields ${\cal A}^a_i({\bf x})$) is
formally the same as 3-dimensional Euclidean QCD with dimensionfull
coupling constant $g_3=g\sqrt{T}$.
This theory is ill behaved in perturbation theory
(the IR divergences in Fig. 1 may be seen as an illustration
of such a bad IR behaviour \cite{MLB}),
but it is generally believed~\cite{KN96} to generate a dynamical mass gap
$\propto g^2_3=g^2 T$, which is the only mass scale in the problem.
This is the celebrated ``magnetic mass'' alluded to in Sec. 3.

\begin{figure}
\protect \epsfxsize=12.cm{\centerline{\epsfbox{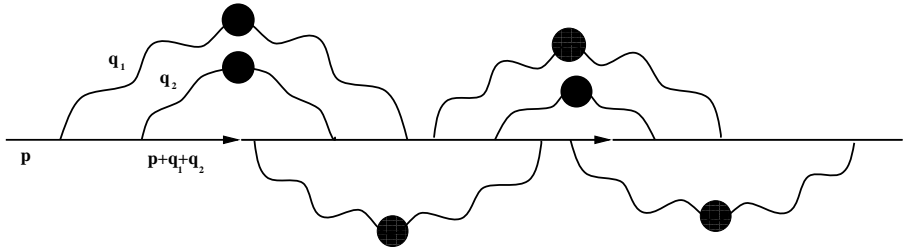}}}
	 \caption{A generic self-energy correction which yields IR divergences
in perturbation theory. All the diagrams of this kind are resummed in the 
non-perturbative calculation of the fermion propagator based on Eq.~(\ref{Z}).}
\label{NDAMP}
\end{figure}
As a second application, consider the large-time behaviour of the electron,
or quark, propagator, which we have seen to be ill defined in perturbation theory
 (cf. Sec. 3).
Fermions have not been yet included in the effective theory, but it is easy to
compute the fermion propagator in a soft background field $A^i$
(in the eikonal approximation \cite{lifetime}), and then average over
the thermal fluctuations of the background field as shown in Eq.~(\ref{Z}).
In QED, all these calculations can be done explicitly (the corresponding functional
integral is Gaussian), with the striking result that, at very large times,
the electron propagator shows a {\it non-exponential} decay  \cite{lifetime}:
$S(t)\propto \exp\{-\alpha T \, t\, \ln(m_D t)\}$, with $\alpha=e^2/4\pi$.
In terms of diagrams, this calculation corresponds to a resummation of all
the ``quenched'' self-energy corrections illustrated in Fig. 3.

In QCD, the corresponding calculation can be performed only 
numerically, which requires a lattice implementation of the effective theory.
More generally, such an implementation would allow for systematic studies
of the real-time non-perturbative dynamics in hot gauge theories.


\section*{References}
\begin{thebibliography}{99}

\bibitem{BIO96}
 J.P. Blaizot, J.-Y. Ollitrault and E.~Iancu, 
in {\it Quark-Gluon Plasma 2}, ed. R.C. Hwa (World Scientific, 
Singapore, 1996).

\bibitem{MLB}
M. Le Bellac, {\it Recent Developments in Finite Temperature Quantum
 Field Theories}, (Cambridge University Press, Cambridge, 1996).

\bibitem{AZ}
P. Arnold and C. Zhai, \PRD{51}{1995}{1906}; 
B. Kastening and C. Zhai,  {\it ibid.} {\bf D52}, 7232 (1995);
E. Braaten and A. Nieto,   {\it ibid.} {\bf D53}, 3421 (1996).

\bibitem{DeTar}
C. DeTar, in {\it Quark-Gluon Plasma 2}.

\bibitem{Kajantie}
K. Kajantie, M. Laine, K.~Rummukainen and M.~Shaposhnikov, \NPB{458}{1996}{90};
 {\it ibid.} \PRL{77}{1996}{2887}.


\bibitem{Hatsuda}
B. Kastening, \PRD{56}{1997}{8107}; T. Hatsuda,
  {\it ibid.} {\bf D56}, 8111 (1997).


\bibitem{Shapo}
{ For a recent review, see} 
V.A. Rubakov and M. Shaposhnikov, hep-ph/9603208,
 Usp. Fiz. Nauk. {\bf 166}, 493 (1996).

\bibitem{Kelly}
P.F. Kelly  {\it et al.}, \PRL{72}{1994}{3461}.

\bibitem{Yaffe} P. Arnold, D. Son and L. Yaffe,\PRD{55}{1997}{6264};
 P. Arnold, {\it ibid.} {\bf D55}, 7781 (1997).

\bibitem{Pisarski93}
V.V Lebedev and A.V. Smilga, \PLB{253}{1991}{231};
R.D. Pisarski, \PRD{47}{1993}{5589}.

\bibitem{Iancu}
E. Iancu,  hep-ph/9710543.

\bibitem{AmbK} J. Ambj{\o}rn and A. Krasnitz, \PLB{362}{1995}{97}.

\bibitem{Hu}
G.D. Moore,  C. Hu and B. M{\"u}ller, hep-ph/9710436.

\bibitem{McLerran} D. B{\"o}deker, L. McLerran and A. Smilga, Phys.
        Rev. {\bf D52}, 4675 (1995).

\bibitem{Nair}
V.P.~Nair, \PRD{48}{1993}{3432}; {\it ibid.} {\bf D50}, 4201 (1994).

\bibitem{Braaten}
E. Braaten, \PRL{74}{1995}{2164}.

\bibitem{Karsch}
F. Karsch {\it et al.}, \PLB{390}{1997}{275}.

\bibitem{Kajantie2}
K. Kajantie {\it et al.}, \PRL{79}{1997}{3130}.

\bibitem{KN96}
W. Buchm\"uller and O. Philipsen, \NPB{443}{1995}{47};
D. Karabali and V.P. Nair, \PLB{379}{1996}{141};
R. Jackiw and S.-Y. Pi, \PLB{403}{1997}{297}.

\bibitem{lifetime}
J.P. Blaizot and E. Iancu,  \PRL{76}{1996}{3080}; 
{\it ibid.} \PRD{56}{1997}{7877}.

\end{thebibliography}
\end{document}